\documentclass{raa}

\usepackage[varg]{txfonts}

\usepackage{natbib}
\usepackage{graphicx}
\usepackage{epstopdf}
\usepackage{amssymb}
\usepackage{float}

\bibpunct{(}{)}{;}{a}{}{,}

\begin{document}

\title{Determination of the Magnetic Fields of Magellanic X-Ray Pulsars}

   \volnopage{Vol.0 (200x) No.0, 000--000}      
   \setcounter{page}{1}          

\author{
Dimitris M. Christodoulou\inst{1,2},
Silas G. T. Laycock\inst{1,3},
Jun Yang\inst{1,3},
\and 
Samuel Fingerman\inst{1,3}
}

\institute{
Lowell Center for Space Science and Technology, University of Massachusetts Lowell, Lowell, MA, 01854, USA.\\
\and
Department of Mathematical Sciences, University of Massachusetts Lowell, Lowell, MA, 01854, USA. E-mail: dimitris\_christodoulou@uml.edu \\
\and
Department of Physics \& Applied Physics, University of Massachusetts Lowell, Lowell, MA, 01854, USA. E-mail: silas\_laycock@uml.edu, jun\_yang@uml.edu, fingerman.samuel@gmail.com \\
}

\date{Received~~2016 month day; accepted~~2016~~month day}

\def\gsim{\mathrel{\raise.5ex\hbox{$>$}\mkern-14mu
             \lower0.6ex\hbox{$\sim$}}}

\def\lsim{\mathrel{\raise.3ex\hbox{$<$}\mkern-14mu
             \lower0.6ex\hbox{$\sim$}}}

\abstract{
The 80 high-mass X-ray binary (HMXB) pulsars that are known to reside in the Magellanic Clouds (MCs) have been observed by the {\it XMM-Newton} and {\it Chandra} X-ray telescopes on a regular basis for 15 years, 
and the {\it XMM-Newton} and {\it Chandra} archives contain nearly complete information about the duty cycles of the sources with spin periods $P_S<100$~s. We have rerprocessed the archival data from both observatories and we combined the output products with all the published observations of 31 MC pulsars with $P_S < 100$~s in an attempt to investigate the faintest X-ray emission states of these objects that occur when accretion to the polar caps proceeds at the smallest possible rates. These states determine the so-called propeller lines of the accreting pulsars and yield information about the magnitudes of their surface magnetic fields. We have found that the faintest states of the pulsars segregate into five discrete groups which obey to a high degree of accuracy the theoretical relation between spin period and X-ray luminosity. So the entire population of these pulsars can be described by just five propeller lines and the five corresponding magnetic moments
($0.29, 0.53, 1.2, 2.9,$ and 7.3, in units of~$10^{30}$~G~cm$^3$).
}

\authorrunning{Christodoulou et al.}
\titlerunning{Magnetic Fields of Magellanic X-Ray Pulsars}

\maketitle


\section{Introduction and Motivation}\label{intro}

The {\it XMM-Newton} and {\it Chandra} X-ray observatories have monitored the
Magellanic Clouds (MCs) on a regular basis in the years 2000-2014 and their archives
contain a wealth of information about the 80 X-ray sources that were identified to be 
members of high-mass X-ray binaries (HMXBs). We have re-analyzed all Magellanic data 
from both archives and we created a new pipeline of products that delineate the 
physical properties of HMXBs and their accreting pulsars. The details of our analysis,
the resulting library, and the public release of the raw and processed data are described in the
works of \cite{yang17} and \cite{chr16}. In this investigation, we extend the work of
\cite{chr16} who focused on the faintest X-ray observations in order to map out the lowest
states of pulsed X-ray emission that occur when accretion to the polar caps proceeds
at the smallest possible rates. Of the entire population of X-ray pulsars, only eight of them 
(lowest row in Table~\ref{t1}) were observed in such states and their lowest-luminosity observations appeared
to define the absolutely lowest propeller line in the spin period-luminosity ($P_S$-$L_X$) diagram 
of the MC pulsars. The best-fit line through these eight points turned out to be in excellent
agreement with the theoretical propeller line, as this was derived by \cite{ste86}, and it
yielded a canonical magnitude for the surface magnetic field of these pulsars of 
$B=3\times 10^{11}$~G (or a magnetic moment of $\mu = 3\times 10^{29}$~G~cm$^3$).

\begin{figure}
\includegraphics[scale=1]{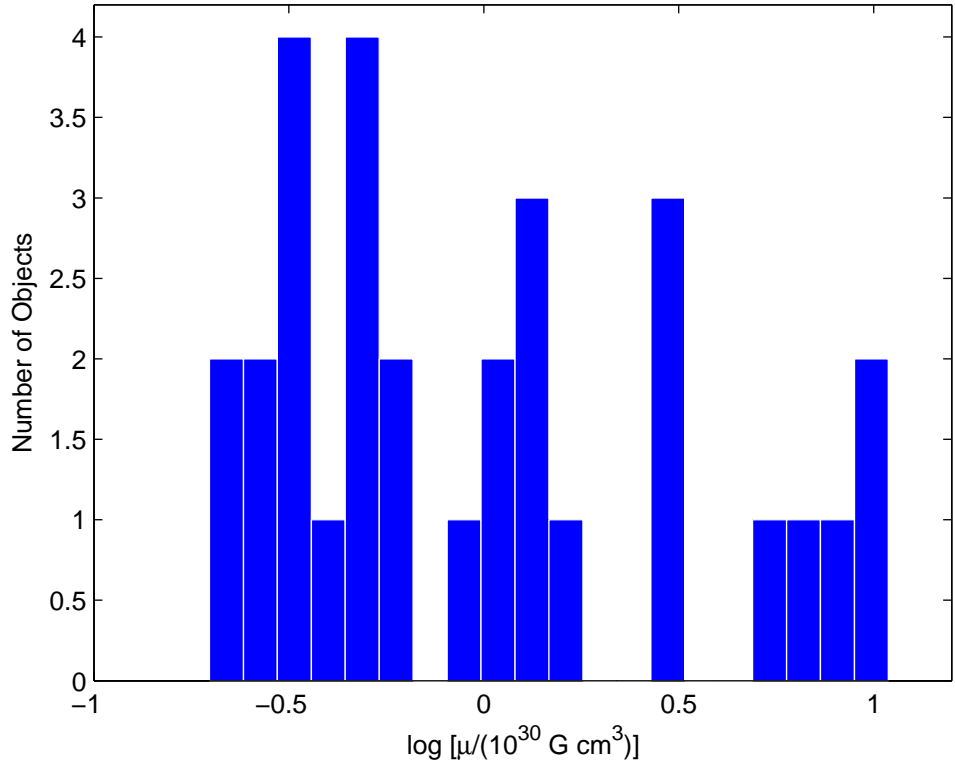}
\caption{Histogram of $\log\mu$ values obtained from equation~(\ref{stella2}) for the Magellanic pulsars with spin periods $P_S<100$~s. The data were presented in \cite{chr16}. The raw data was distributed in 20 equally spaced bins between the values $-0.704$ and 1.04. Five peaks are readily visible and the pulsars that cluster around each peak are listed in Table~\ref{t1}.
}
\label{fig000}
\end{figure}

%
\begin{table}

\caption{Discrete Propeller Lines in the $P_S$-$L_X$ Diagram}
\label{t1}
\centering
\begin{tabular}{c c c}
\hline\hline
Propeller & Number          & Contributing\\
Line        & of MC Pulsars  & MC Pulsars\\
\hline

Highest & 5 & SXP: 4.78, 6.85, 74.7 and LXP: 28.8, 61.6 \\	
Fourth & 3 & SXP: 5.05, 7.78, 59.0 \\	
Third   & 7 & SXP: 2.37, 8.02, 25.5, 31.0, 46.6, 82.4 and LXP: 4.40 \\		
Second& 7 & SXP: 0.72, 7.92, 9.13, 11.6, 11.87, 15.3 and CXOU J010043.1-721134\\
Lowest & 8 & SXP: 3.34, 6.88, 8.88, 18.3, 22.1 and LXP: 0.07, 4.10, 8.04 \\
								
\hline
\end{tabular}
\end{table}

\begin{figure}
\includegraphics[scale=1]{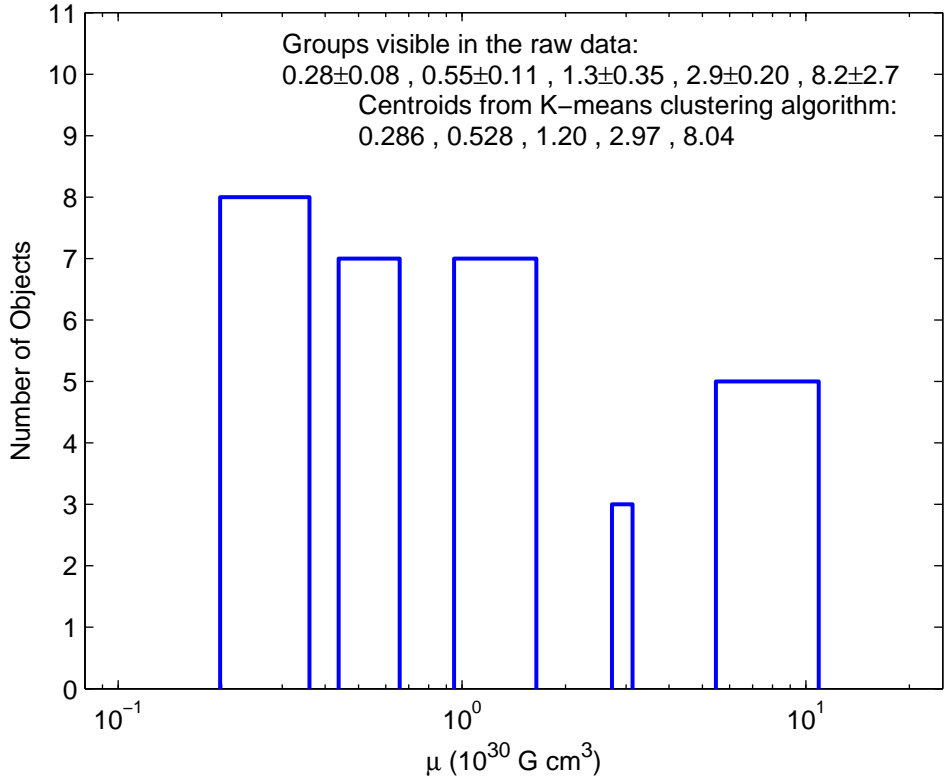}
\caption{Histogram of the magnetic moments $\mu$ obtained from equation~(\ref{stella2}) for the sources listed in Table~\ref{t1}.
The data were presented in \cite{chr16}. The gaps between bars are very large, comparable to or larger than the half-widths of the bars. A K-means clustering algorithm minimizing Euclidean squared distances also confirms the observed clusters and shows that there is no overlap between clusters (see Figure~\ref{fig00}). It is clear that the magnetic fields of the Magellanic pulsars are segregated into five distinct groups, as specified in the legend and in Table~\ref{t1}.
}
\label{fig0}
\end{figure}

\begin{figure}
\includegraphics[scale=1]{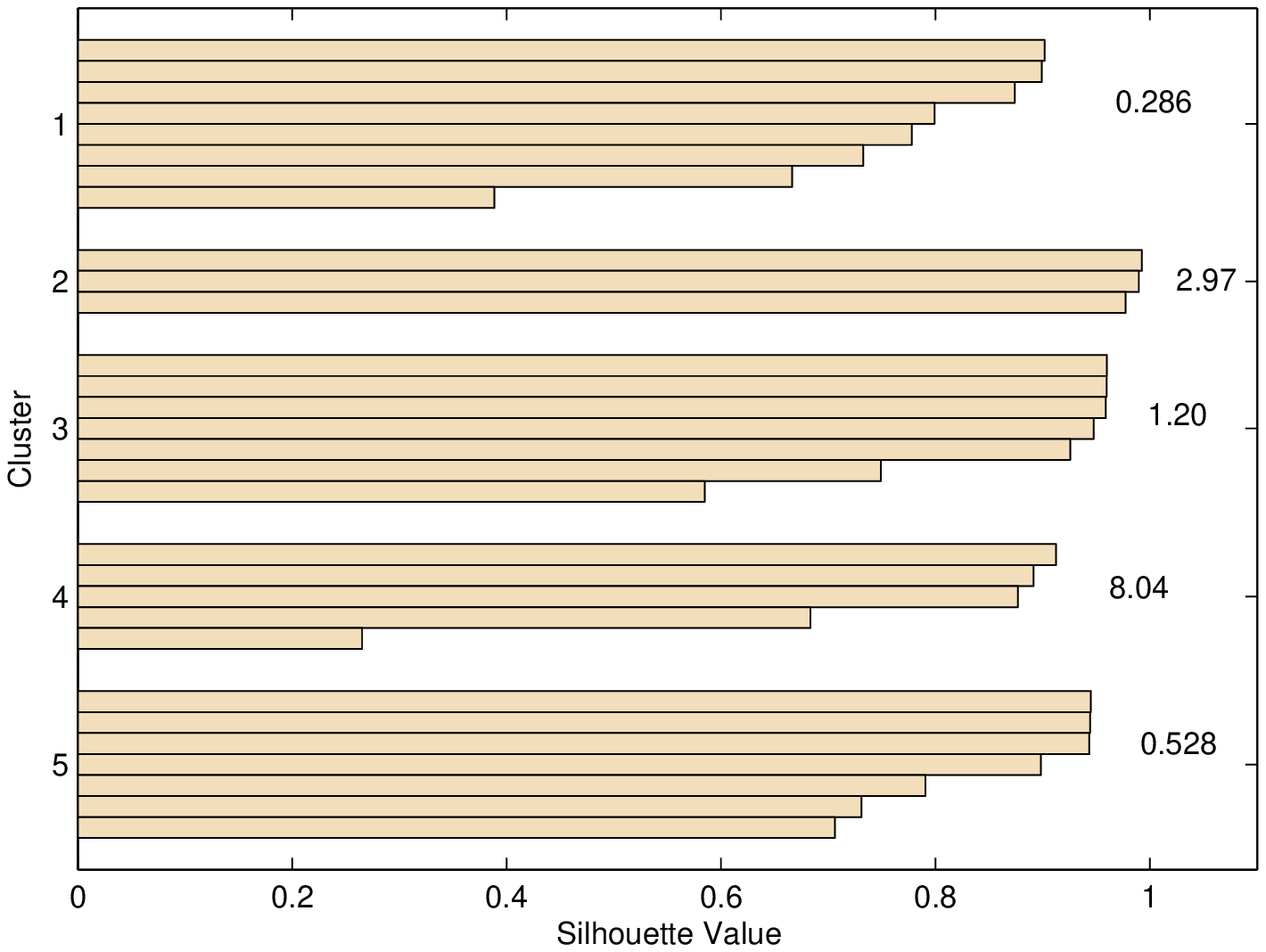}
\caption{The silhouette diagram from the K-means clustering algorithm that minimizes Euclidean squared distances. The algorithm finds five clusters of pulsars (see Table~\ref{t1}). The silhouette values are very large ($\geq0.6$) for nearly all of the pulsars, indicating that the clusters are well-separated. The centroids of the clusters are also shown, and they are in excellent agreement with the group centers determined from the raw data (see Figure~\ref{fig0}). 
}
\label{fig00}
\end{figure}

The rest of the X-ray sources did not appear to reach down to the lowest propeller line
despite the fact that the duty cycles of short-period pulsars (with $P_S<100$~s) were covered very well by both observatories over a
period of 15 years. This implies that these pulsars have stronger magnetic fields and their
faintest states are lying significantly higher than the lowest propeller line determined by \cite{chr16}.
In this work,
we adopt again the assumption that the minimum X-ray luminosities of these sources will not
be randomly distributed in the $P_S$-$L_X$ diagram; instead, they will cluster
in groups, each characterized by a typical higher value of their magnetic fields, owing to the similarities of these accreting pulsars in their structures and in their evolutionary paths. We searched for such progressively higher values of the magnetic fields in the lowest-power observations of the X-ray sources listed in Tables~1-3 of \cite{chr16}. 
A histogram of $\log\mu$ values distributed in 20 equally spaced bins is shown in Figure~\ref{fig000}. Five peaks are visible in the raw data, indicating that it is worth pursuing a formal clustering analysis. The pulsars that cluster around each peak are shown in Table~\ref{t1}.

\subsection{Clustering Analysis}

The five groups of magnetic moments and their boundaries are represented in Figure~\ref{fig0} using bins centered around the peaks in Fig.~\ref{fig000}. One can specify these bins in the raw data because of the uncharacteristically large gaps between groups. Furthermore, an investigation of clustering in the data using the K-means algorithm \citep{seb84,spa85} and minimizing Euclidean squared distances confirms the dense clustering of these five groups. The centroids of the clusters are shown in the legend of Figure~\ref{fig0}, and they are in excellent agreement with the groups found by visual inspection of the data. 

We also constructed the ``silhouette diagram'' of the data \citep{rou87,kau90}, as shown in Figure~\ref{fig00}. This diagram shows the significance of clustering. The silhouette values of nearly all of the pulsars are very large ($\geq 0.6$), indicating that the clusters are well-separated with no overlap at all; and only two members (in clusters 1 and 4, respectively) are located at the outskirts of their clusters that are otherwise very dense, just like clusters 2, 3, and 5. These two outliers have positive silhouette values, thus they cannot be moved to the nearest neighboring cluster.

The two lowest clusters in $\mu$ shown in Figure~\ref{fig0} are separated by a gap of only 0.08, so one might think that they may be merged into one group. We ran the K-means algorithm seeking only 4 clusters, but in this case the quality of the results was degraded. The silhouette diagram showed 7 outliers of which 4 were in the merged cluster with centroid $0.37\pm 0.17$. 

The dense clustering seen in the above figures is not by itself evidence that the Magellanic pulsars are segregated into five well-defined groups. If one chooses 30 $\log\mu$ values in ($-1, 1$) from a uniform random distribution and distributes them into 20 bins, some groups are bound to appear by chance. But when the known spin periods are also randomly assigned to these points, the resulting $P_S$-$L_X$ diagrams do not resemble the distribution of the observed $L_X$ values versus spin period. In particular, the numerical experiments do not reproduce the lowest propeller line found for $\mu = 3\times 10^{29}$~G~cm$^3$ or the theoretical relation \citep{ste86} within each group (see \S~\ref{outline} below).

To rectify this problem, we limited the $L_X$ parameter space between the lowest propeller line \citep{chr16} and the canonical Eddington line ($L_{Edd}=1.8\times 10^{38}$~erg~s$^{-1}$). In that case, the groups at high spin periods were wiped out because of the large ranges in $L_X$ values that occur at higher spin periods. Therefore, these random distributions also do not resemble the demarkation shown in Table~\ref{t1}, especially in the cases Highest to Third.

\subsection{Outline}\label{outline}

Irrespective of clustering arguments, the above five clusters of pulsars will be physically meaningful only if, in addition, they obey the theoretical relation
for the propeller line \citep{ste86}. In \S~\ref{lines}, we describe our investigation of
such additional propeller lines and their comparison with theoretical predictions. In \S~\ref{conc}, we summarize and discuss our results.

\section{Propeller Lines and Magnetic Fields in the Magellanic Clouds}\label{lines}

\begin{figure}
\includegraphics[scale=1]{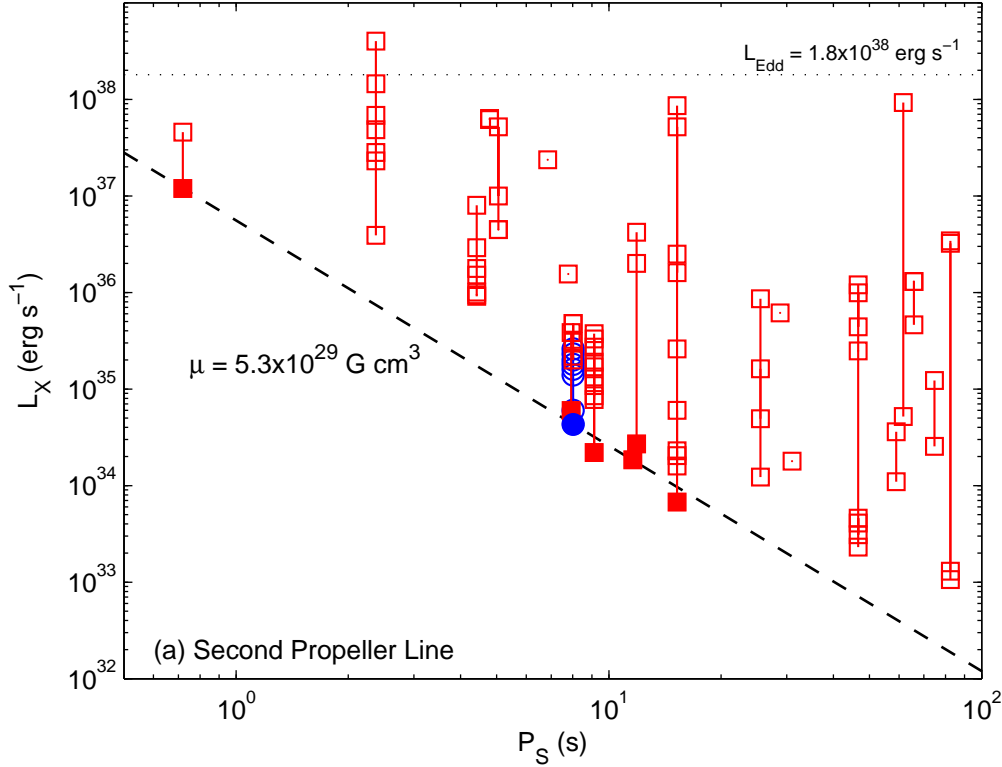}
\caption{$L_X$ vs. $P_S$ diagram for the X-ray sources listed in Table~\ref{t1}.
The data were presented in \cite{chr16} and the observations of the sources
that defined the lowest propeller line have been removed. 
In each iteration from (a) to (d), the observations of the pulsars defining lower-lying propeller lines are removed. Unfilled symbols represent individual observations. Dots mark multiple observations with nearly identical results. Filled symbols
in each panel represent the pulsars that define the corresponding propeller line
listed in Tables~\ref{t1} and~\ref{t2}. 
Blue circles in panel (a) represent observations of the single pulsar CXOU J010043.1-721134.
The dashed line is the theoretical propeller line with a slope of $-7/3$ that implies the
magnetic-moment value shown in each panel.
The dotted line is the Eddington luminosity for a 1.4~$M_\odot$ pulsar.
}
\label{fig1}
\end{figure}

We process the data (from {\it XMM-Newton}, {\it Chandra}, and the published literature) presented in \cite{chr16} as follows: First we remove the upper limits in cases of no detection and the few extremely faint (and uncertain) detections that fell below the lowest propeller line which may represent
unpulsed magnetospheric emission \citep{cam95,cor96,cam97}, if they are real. Then we follow an iterative process: we remove all the observations of the sources that defined the lowest propeller line and we consider the faintest observations that appear to define the next higher propeller line according to Figures~\ref{fig0} and~\ref{fig00}. In each subsequent step, we remove all the observations of the sources that define
lower-lying propeller lines and we consider the next cluster of pulsars. The iteration continues until all
the data are exhausted. This process results in the five dense groups of pulsars discussed above (Table~\ref{t1} and Figure~\ref{fig1}).

\begin{figure}
\includegraphics[scale=1]{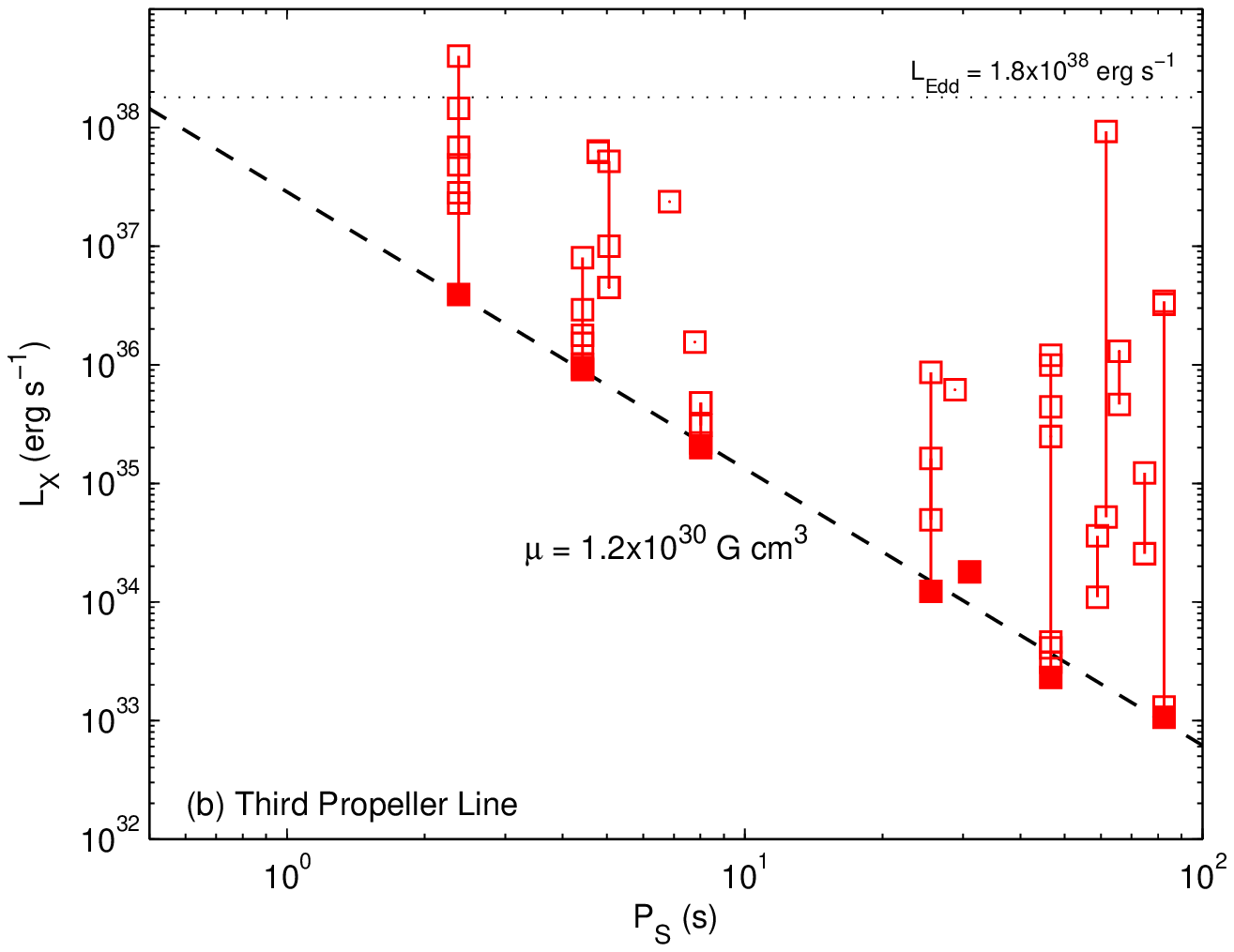}
\end{figure}
\begin{figure}
\includegraphics[scale=1]{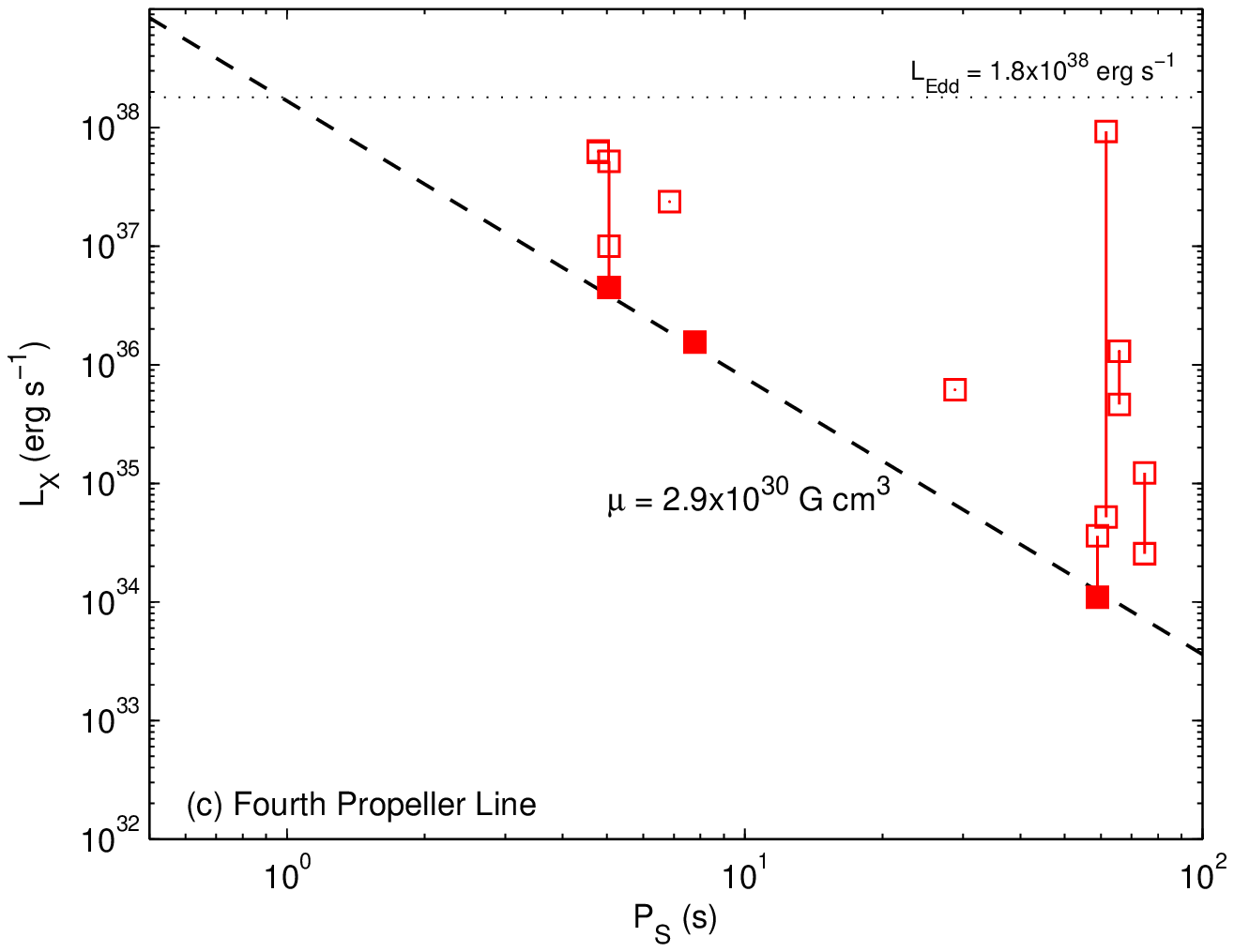}
\end{figure}
\begin{figure}
\includegraphics[scale=1]{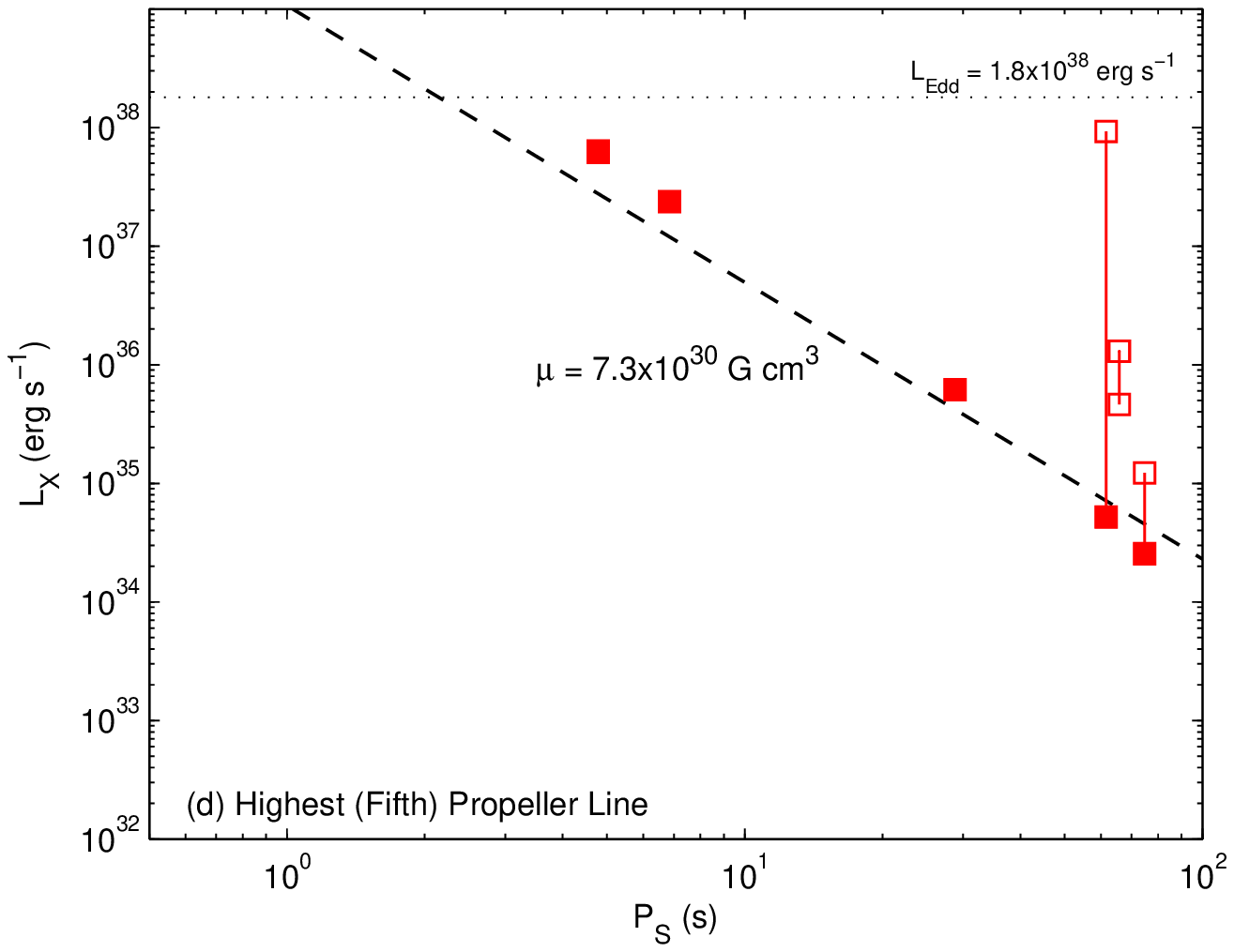}
\end{figure}

For each of the five groups, we carry out a linear regression of the faintest observations in the 
$\log P_S$-$\log L_{X,min}$ diagram in order to determine how close each best-fitted slope is to the theoretical 
value of $-7/3$ (Table~\ref{t2}). The precise theoretical propeller line is obtained from the equation given by
\cite{ste86} and for the canonical pulsar parameters $M=1.4~M_\odot$ and $R=10$~km, viz.
\begin{equation}
L_{X,min} \approx 2\times 10^{37} \left(\frac{\mu}{10^{30}~{\rm G~cm^3}}\right)^2 
\left(\frac{P_S}{1~{\rm s}}\right)^{-7/3}{\rm erg~s^{-1}}\, .
\label{stella2}
\end{equation}
This equation contains only one free parameter, the magnetic moment of the pulsar 
\citep[$\mu\equiv B R^3$, as was defined by][]{ste86}
assuming a dipolar magnetic field of magnitude $B$ on its surface. We note that the magnetic moments determined from the \cite{ste86} equation are subject to variations of no more than 54\% if large noncanonical values are used and that most of the uncertainty comes from the pulsar radius: when we adopt $\Delta M=0.6 M_\odot$ and $\Delta R=8$~km \citep{lat01} for the variation of $\mu\sim M^{1/3}R^{1/2}$, we find that $\Delta\mu /\mu = 54$\% in which only 14\% is contributed by the variation of the mass.

%
\begin{table}

\caption{Linear Regressions for the 5 Propeller Lines}
\label{t2}
\centering
\begin{tabular}{c c c c c c c}
\hline\hline
Propeller & Number & Slope  & Deviation & Intercept & $p$-value & Correlation\\
Line        & of MC Pulsars & ($\pm 1\sigma$ Error)  & from $-7/3$ Slope & ($\pm 1\sigma$ Error) & & ($r^2$)\\
\hline

Highest  & 5 & $-2.795$ ($\pm 0.089$) & $-19.80\%$ & 39.725 ($\pm 0.115$) & 0.0203 & 0.997\\
Fourth   & 3 & $-2.445$ ($\pm 0.001$) & $-4.80\%$ & 38.371 ($\pm 0.001$)   & 0.0002 & 1.000 \\
Third     & 7 & $-2.334$ ($\pm 0.115$) & $-0.04\%$ & 37.458 ($\pm 0.142$)   & 0.0313 & 0.988 \\
Second  & 7 & $-2.347$ ($\pm 0.131$) & $-0.60\%$ & 36.758 ($\pm 0.116$)   & 0.0355 & 0.985 \\
Lowest  & 8 & $-2.322$ ($\pm 0.092$) & $+0.47\%$ & 36.207 ($\pm 0.088$)  & 0.0252 & 0.991\\

\hline

\end{tabular}

\end{table}

Table~\ref{t2} shows that the four lower best-fitted lines are in excellent agreement with the corresponding theoretical propeller lines (the null hypothesis is rejected at the 95\% confidence level). Only for the fifth and highest line, for which we used the last five remaining objects, is the error in the slope substantial (about $20\%$)\footnote{Even in this case, when we include the published observations of pulsars with spin periods $P_S > 100$~s, we find only two more propeller states \citep[SXP131, $L_X = 2.7\times 10^{34}$~erg~s$^{-1}$; SXP342, $L_X = 3.8\times 10^{33}$~erg~s$^{-1}$;][]{lay10}, and the slope ($-2.358\pm 0.179$) returns gracefully to a value that differs from $-7/3$ by only 1\%; then the intercept ($39.244\pm 0.290$) gives a magnetic field of $B=9.4$~TG ($r^2 = 0.972, p=0.048$) for the highest propeller line.}, although its $p$-value is still significant.
To make up for this deviation in slope, we produced another linear fit to the data
in which we fixed the intercepts to the values that yield a slope of $-7/3$ exactly. These results
are shown in Table~\ref{t3}. The correlation coefficients ($r^2$) indicate that these fits are also of high quality. The $p$-values of these fits are extremely small, but they are meaningless because of the imposed constraint.
However, even the highest propeller line with a constrained slope of $-7/3$ is an acceptable fit to the few remaining data points ($r^2 = 0.966$).

%
\begin{table}
\caption{Linear Fits Imposing a Slope of $-7/3$ to the Data}
\label{t3}
\centering
\begin{tabular}{c c c c c}
\hline\hline

Propeller & Intercept & Correlation & $\mu$ & $B$ \\
Line &  (for slope $\equiv -7/3$) & ($r^2$) & (G~cm$^3$)  & (TG)\\

\hline

Highest  &  39.029   & 0.966  &   $7.3\times 10^{30}$    &   7.3  \\
Fourth   &   38.224  & 0.998  &  $2.9\times 10^{30}$     &  2.9  \\
Third     &   37.456  & 0.988   &  $1.2\times 10^{30}$    &   1.2  \\
Second  &  36.744   & 0.985   &   $5.3\times 10^{29}$    &   0.53  \\
Lowest  &   36.224  & 0.991    &  $2.9\times 10^{29}$   &    0.29 \\

\hline

\end{tabular}

\end{table}

The magnetic moments listed in Table~\ref{t3} were then determined from the intercepts of the fits
using equation~(\ref{stella2}). The corresponding magnetic-field magnitudes were also 
determined using
the definition $\mu\equiv B R^3$ and the canonical pulsar radius of $R=10$~km.
The range of magnetic fields ($B\approx 0.3-7.3$~TG) is consistent with the values
quoted in the literature for many Galactic and extragalactic HMXBs; those obtained by applying accretion theory \citep{ste86,ste94,cam97,cor97,gal08,bac14}; and those obtained by {\it NuSTAR} observations of cyclotron resonance features in the 14-32~keV energy range \citep{ten14,bri16}.

\section{Summary and Discussion}\label{conc}

We have processed the Magellanic HMXB data from the {\it XMM-Newton} and the {\it Chandra} archives and from the published literature that were listed in \cite{chr16}
who determined the lowest propeller line in the spin period-luminosity ($P_S$-$L_X$) diagram on which
accretion proceeds at the smallest possible rates and the X-ray emission is pulsed.
We used the K-means clustering algorithm \citep{seb84,spa85} followed by an iterative process in order to search for additional propeller lines and we found
a total of five such lines for pulsars with $P_S<100$~s (Tables~\ref{t1} and~\ref{t2}; Figures~\ref{fig0} and~\ref{fig1}). The linear regressions 
of four data sets produced best-fitted lines of high quality that are additionally in excellent agreement
with the theoretical $P_S$-$L_X$ relation of \cite{ste86}. The fifth data set that
resulted in the highest propeller line can also be fitted with the theoretical relation
to a satisfactory degree (Table~\ref{t3}).

The linear fits of the theoretical relation (equation~[\ref{stella2}]) were used to determine the magnetic fields of the five groups of pulsars.
The entire population of MC pulsars is described by the propeller lines shown in Table~\ref{t3} 
and the corresponding magnetic-field magnitudes
($0.29, 0.53, 1.2, 2.9,$ and 7.3, in units of~TG).
These discrete values may come as a surprise because there is no {\it a priori} reason for the low-luminosity X-ray observations of all of these objects
to line up on precise straight lines such as those shown in Tables~\ref{t2} and~\ref{t3} and in Figure~\ref{fig1}, unless of course several different factors concur:  
\begin{itemize}
\item[(a)] the X-ray duty cycles of the pulsars were very well monitored during all of these 15 years; 
\item[(b)] the observations did manage to probe the lowest levels of pulsed X-ray emission from each pulsar; and
\item[(c)] the physical characteristics of the X-ray emitting regions, the accretion processes, and the pulsars themselves are very similar for the entire population \citep[as was also found by][]{coe10}.
\end{itemize}

It will be interesting to see if these five groups of MC pulsars (Table~\ref{t1}) do indeed provide robust classes based on their surface magnetic fields, as more of their physical properties will be obtained in the following years that will produce more classifications of these objects. We believe that our classification of Magellanic pulsars will prove very useful to such future investigations.

\begin{acknowledgements}
This work was supported by NASA grant NNX14-AF77G.
\end{acknowledgements}



\label{lastpage}

\end{document}